# The Binding Energy of Triplet Excitons in Non-Fullerene Acceptors: The Effects of Fluorination and Chlorination

J. P. A. Souza,[a] L. Benatto,[a] G. Candiotto,[b] L. S. Roman[a] and M. Koehler[a]

[a] Department of Physics, Federal University of Paraná, 81531-980, Curitiba-PR, Brazil

[b] Institute of Chemistry, Federal University of Rio de Janeiro, 21941-909, Rio de Janeiro-RJ, Brazil

Corresponding authors: joaojp@fisica.ufpr.br, koehler@fisica.ufpr.br

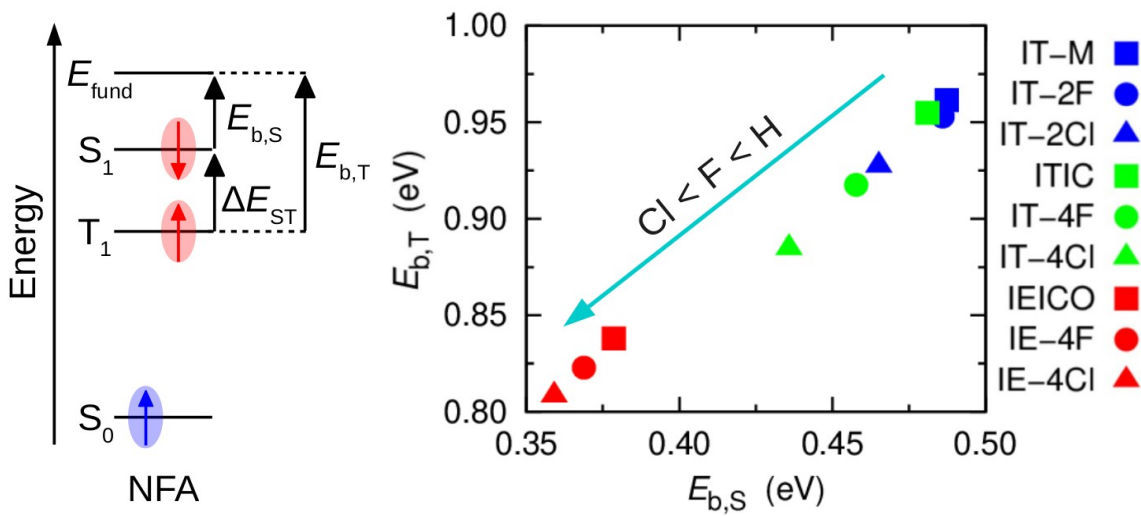

**TOC Graph**

Abstract

One strategy to improve the photovoltaic properties of non-fullerene acceptors (NFAs) is the rational fluorination or chlorination of those molecules. Although this modification improves important acceptor properties, little is known about the effects on the triplet states. Here, we combine the polarizable continuum model with optimally tuned range-separated hybrid functional to investigate this issue. We find that fluorination or chlorination of NFAs decreases the degree of HOMO-LUMO overlap along these molecules. Consequently, the energy gap between $T_1$ and $S_1$ states, $\Delta E_{ST}=E_{S1}-E_{T1}$, also decreases. This effect simultaneously enhances the generation of triplet excitons and reduce the binding energy of the triplet excitons ($E_{b,T}$) which favor their dissociation into free charges. Interestingly, although Cl has a lower electronegativity than F, the chlorination is more effective to reduce $\Delta E_{ST}$. Since chlorination of NFAs is easier than fluorination, Cl substitution can be a useful approach to enhance solar energy harvesting using triplet excitons.

Over the years, the increased performance of organic solar cells (OSCs) has been directly linked to the development of more efficient organic semiconductor materials.[1,2] Generally, two types of solution-processed organic semiconductors are employed in the active layer of OSCs: (i) electron donor polymer (D) and (ii) electron acceptor molecule (A). The combination of these two materials forms a D/A heterojunction.[3,4] Currently, the performance of laboratory-scale OSCs are more than 18%[5–7] which raises the potential for energy production using low cost devices processed by solution printing or coating techniques.[8,9] In addition, the semi-transparency and flexibility characteristics of OSCs enable its application in an innovative way in comparison with inorganic solar cells.[10]

However, photoexcitation of organic semiconductors generates excitons that are coulombically bound electron–hole pairs.[11] The excitons formed in D or A can diffuse up to the D/A interface where they are more easily dissociated, thus generating free charges.[12–14] The free charges are then transported to their respective electrodes, producing the device photocurrent response.[15]

In recent years, more efficient OSCs have been produced with non-fullerene acceptors (NFAs).[16] Compared with traditional fullerene acceptors (FAs), NFAs have enhanced light absorption and tunable energy levels of the frontier molecular orbitals.[17,18] In addition, NFAs have lower exciton binding energy of the singlet state,[19] allowing the generation of free charges with lower driving force (energy difference between the local excited state and the charge transfer state).[20] A lower driving force is important to increase the device's open circuit voltage.[21–23] There are also reports of OSCs based on NFAs that showed very interesting characteristics such as ultrafast charge transfer,[24,25] low charge trapping[26] and high morphological stability.[27] Despite these interesting characteristics of NFAs, further improvements have been sought.[28] Among them we can highlight the use of the two types of excited states, *i.e.*, singlets ($S_1, ..., S_n$) and triplets ($T_1, ..., T_n$), in the photovoltaic process.[29,30]

Direct photon absorption in NFAs generates singlets excitons. Yet triplet excitons can be generated by intersystem crossing (ISC) and singlet fission.[31] Triplet excitons are interesting because they have longer lifetimes than singlet excitons due to the forbidden nature of their recombination.[32] In principle, this property can lead to longer diffusion lengths, which would tend to decrease the morphological constrains associated to the sizes of the donors (acceptors) domains in bulk heterojunctions.[33,34] On the other hand, triplet excitons have a higher binding energy compared to singlet excitons due to the attractive exchange interaction of the same spin orientation. This property makes them more difficult to dissociate. Importantly, when the energy gap between T$_1$ and S$_1$ decreases, $\Delta E_{ST}=E_{S1}-E_{T1}$, the exciton binding energy of the triplet exciton ($E_{b,T}$) approaches the binding energy of the singlet exciton ($E_{b,S}$) which would help the dissociation of the triplet excitons. In addition, the reduction of $\Delta E_{ST}$ enhances the intersystem crossing (ISC) from S$_1$ to T$_1$ which favors the generation of T$_1$ excitons.[35] Some of those characteristics of triplet and singlet excitons were experimentally observed in oligoacenes by Hummer *et al.*[36] and theoretically reproduced with good precision by Hu *et al.*[37]. In the latter study, the calculations were performed combining the polarizable continuum model (PCM) and optimally tuned range-separated (RS) hybrid functional. From a theoretical point of view, the great importance of considering solid-state polarization effects for the stabilization of singlet and triplet energies was stressed by Chen and coauthors.[38]

The decrease in $\Delta E_{ST}$ is essential to improve the use of triple excitons in the photovoltaic process. Considering a simple two-electron two-state model, it is possible to obtain the relation $\Delta E_{ST} = 2K_{HL}$,[38] where $K_{HL}$ is the electron exchange energy between electron 1 in the HOMO (highest occupied molecular orbital) and electron 2 in the LUMO (lowest unoccupied molecular orbital). $K_{HL}$ depends on the degree of HOMO−LUMO overlap ($\Theta_{H-L}$) and can be reduced with the minimization of $\Theta_{H-L}$. Organic molecules with high twisted D−A structure can effectively separate the HOMO and LUMO ($\Theta_{H-L} \approx 0$) reaching a very small values of $\Delta E_{ST}$.[39] The two main problems of a twisted molecular structure for OSCs applications is the weak oscillator strength (*f*) of the S$_0 \rightarrow$ S$_1$ transition which limits the light absorption and the poor charge carriers mobility over the molecules.[35] Both problems are associated to the reduction in the average conjugated length induced by the twisted structure. In a recent study Qin and coauthors[35] demonstrated that A-D-A-D-A-type NFAs, analogues of Y6,[40] with low twisted conformation (two halves of the molecules share a dihedral angel of approximately 17º) have promising $\Delta E_{ST}$ of approximately 0.35 eV and long lifetime excitons of the order of 50 ns in blend films. In their calculation of $\Delta E_{ST}$, $E_{S1}$ was considered equal to the optical band gap extracted from external quantum efficiency spectra and $E_{T1}$ extracted from the onset of the emission band of the film at low temperature. The effective generation and split of triplet excitons has been reported, contributing to the device photocurrent. Interestingly, the two NFAs considered by Qin presented chlorine and fluorine end-groups.

The promising perspectives for the use of triplet excitons in the photovoltaic process of OSCs[34,41] have motivated us to study more deeply the influence of fluorine and chlorine substitutions on the properties of triplet states generated in NFAs. The fluorination and chlorination processes of organic materials have been widely used in recent years.[42] Both procedures cause similar changes in the morphological and optoelectronic properties of NFAs. A significant improvement in the crystallinity of the formed films was reported, which increases the electronic coupling between molecular orbitals of adjacent molecules and consequently the mobility of electrons.[43] Furthermore, it is reported a shift of the frontier energy levels and a broadening of the absorption spectrum specially with the chlorination.[44] Because the synthesis of chlorinated acceptors is simpler than the synthesis of fluorinated ones, the use of Cl substitutions might be more convenient to reduce large scale production costs for NFA-based OSCs.[45,46]

Here, we will simulate nine NFAs that have a fused acceptor-donor-acceptor linear structure (A-D-A). They can be separated into three groups of three molecules, as shown in Figure 1. Each group contains an unsubstituted molecule, a fluorinated molecule and a chlorinated molecule. The first group consists of the IT-M,[47] IT-2F[48] and IT-2Cl,[44] where the F or Cl atoms replace the methyl attached to the IT-M acceptor units. The second group consists of the ITIC,[49] IT-4F[50] and IT-4Cl,[44] where the F or Cl atoms replace two hydrogens of the ITIC acceptor units. Finally the third group consists of the IEICO,[51] IE-4F[52] and IE-4Cl,[53] where the F or Cl atoms replace two hydrogens of the IEICO acceptor units.

Note that the three groups of NFAs have four side chains linked to the central donor unity. However, group III molecules have two differences from group I and II molecules. The first is a smaller number of aromatic fused rings in the central donor unit, specifically two less fused rings. Note that the fused thiophene present in the central donor unit of the groups I and II molecules make a single bound with this unit in the group III molecules. This modification increases the molecular length of group III molecules. The second difference of group III molecules are two additional side chains linked to thiophenes.

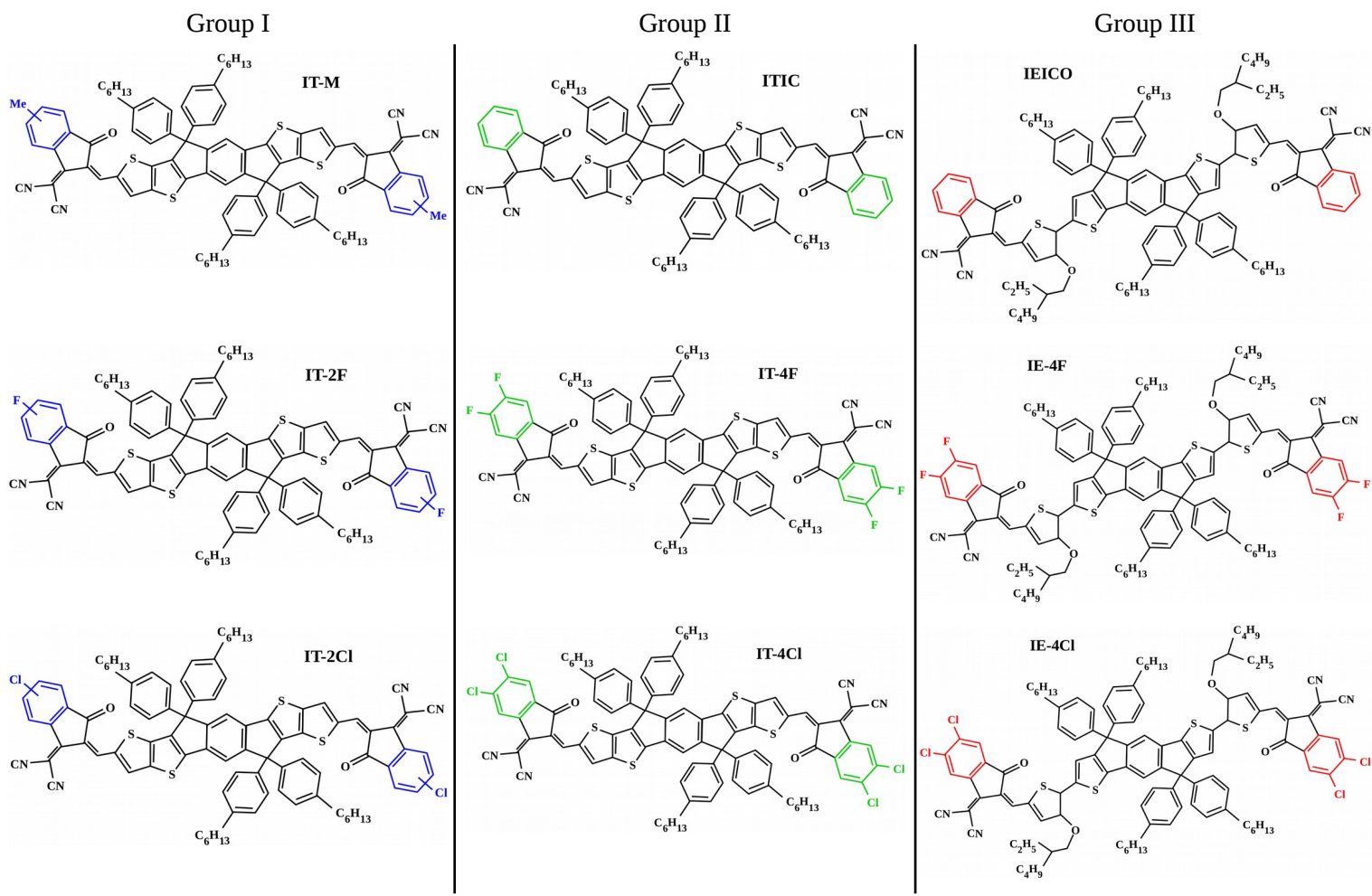

**Figure 1** – Chemical structure of the three groups of NFAs.

We have applied density functional theory (DFT) and time-dependent DFT (TDDFT) calculations to study the key properties of acceptor molecules. All DFT/TDDFT calculations were performed using the Gaussian 16 package.[54]

The first step of our calculations is to optimize the ground state geometry of isolated materials by DFT with the Becke three-parameter Lee-Yang-Parr (B3LYP)[55] hybrid functional and 6-31G(d,p) basis set. To obtain a good description of the molecular energies, the electronic properties of the optimized molecules were calculated *via* DFT/TDDFT using ωB97XD[56] the range-separated hybrid functional with empirical dispersion corrections along 6-31G(d,p) basis set.[57] Besides that, the range-separation parameter ($\omega$) of the functional $\omega$B97XD was optimized following the gap adjustment procedure. In this procedure, $\omega$ is adjusted to minimize $J(\omega)$:[58,59]

$$J(\omega)=|E_{HOMO}(\omega)-IP(\omega)|+|E_{LUMO}(\omega)-EA(\omega)| \qquad (1)$$

where $E_{HOMO}(\omega)$ and $E_{LUMO}(\omega)$ are the energies of the HOMO and LUMO while IP($\omega$) and EA($\omega$) are the vertical first ionization potential and electron affinity of the material, respectively. With the total energies of the cationic ($E_+$), anionic ($E_-$), and neutral ($E_0$) states it is possible to calculate IP = $E_+ - E_0$ and EA = $E_0 - E_-$.[60] This procedure was performed using the polarizable continuum model[61] (PCM) to simulate the molecule in a dielectric medium. It was found that these methods was able to calculate molecular energies that are in better agreement with experimental values.[37,62,63] The dielectric constants of each molecule for the PCM calculation were obtained from ref[64]. We also calculated the atomic charges *via* the electrostatic potential (ESP) method[65–67] to obtain the magnitude of the internal charge transfer (ICT) between the acceptor chemical units and the donor chemical unit of the molecules. The degree of HOMO-LUMO overlap were obtained by the Multiwfn program.[68]

The exciton binding energy can be calculated by subtracting the fundamental energy gap ($E_{fund}$ = IP − EA) from the optical gap.[60] The optical gap can be obtained from TDDFT calculations and corresponds to $E_{S1}$ and $E_{T1}$ for singlet and triplet. Therefore, the binding energy of singlet ($E_{b,S}$) and triplet ($E_{b,T}$) exciton is defined as:

$$E_{b,S}=E_{fund}-E_{S1} \qquad (2)$$

and

$$E_{b,T}=E_{fund}-E_{T1}. \qquad (3)$$

In the sequence we will detail the results obtained using the methods described above. We will pay special attention to the variations in the triplet exciton binding energy with the fluorination and chlorination of the NFAs.

The comparison of the theoretical results of IP, EA and $E_{S1}$ with experimental data obtained by Yang and coauthors[69] can be seen in Table 1 and Figure 2. The three groups of molecules show a similar pattern with respect to the variation of the molecular energies. IP and EA increase with the fluorination or chlorination of the molecules and the opposite occurs for $E_{S1}$. Notice that the variations are greater with the chlorination. We anticipate that this same pattern will be observed below for other electronic properties, *i.e.*, the improvements on key molecular features are higher for chlorination than for fluorination. The correlation coefficient ($R_{sq}$) calculated through the linear adjustment of theoretical and experimental values of IP, EA and $E_{S1}$ resulted in approximately 0.975, 0.933 and 0.943, respectively. The quality of the linear regression to reproduce the data increases as Rsq approaches 1. Thus, there is a reasonable correlation between the theoretical and experimental values of IP, with the theoretical values being on average 0.34 eV lower. For EA the correlation is smaller, where the theoretical values are on average 0.79 eV lower. It is expected a greater deviation from the experimental data for the values of EA since the calculation of this parameter is more sensitive to delocalization error associated with the exchange-correlation functional.[70] Here we point out that comparisons between energies estimated using vertical electronic transitions with energies measured by adiabatic processes (like the CV technique) must always be interpreted with caution. Considering $E_{S1}$, there is a very good correlation between theory and experiment, being the mean deviation of 0.03 eV.

**Table 1** - Optimal Range-Separation Parameter ω (in Bohr$^{-1}$). Theoretical results – in solid with PCM – compared to measurements of IP, EA and $E_{S1}$ energies (in eV).

|  | Molecules | ω | IP | | EA | | $E_{S1}$ | |
|---|---|---|---|---|---|---|---|---|
|  |  |  | Theo. | Exp.[a] | Theo. | Exp.[a] | Theo. | Exp.[a] |
|  | IT-M | 0.0030 | 5.35 | 5.64 | 3.10 | 3.90 | 1.76 | 1.77 |
| I | IT-2F | 0.0030 | 5.42 | 5.69 | 3.20 | 3.99 | 1.72 | 1.73 |
|  | IT-2Cl | 0.0023 | 5.43 | 5.70 | 3.24 | 4.02 | 1.68 | 1.73 |
|  | ITIC | 0.0022 | 5.38 | 5.66 | 3.14 | 3.93 | 1.75 | 1.76 |
| II | IT-4F | 0.0010 | 5.43 | 5.72 | 3.24 | 4.03 | 1.70 | 1.73 |
|  | IT-4Cl | 0.0011 | 5.46 | 5.73 | 3.32 | 4.08 | 1.64 | 1.70 |
|  | IEICO | 0.0006 | 4.95 | 5.38 | 3.12 | 3.90 | 1.53 | 1.45 |
| III | IE-4F | 0.0013 | 5.00 | 5.46 | 3.20 | 4.04 | 1.43 | 1.42 |
|  | IE-4Cl | 0.0001 | 5.03 | 5.49 | 3.28 | 4.08 | 1.40 | 1.39 |

[a] Experimental results of ref[69] in which IP and EA was estimated from electrochemical measurements with cyclic voltammetry (CV) and $E_{S1}$ was estimated from the first maximum absorption wavelength ($\lambda_{max}$) in thin film.

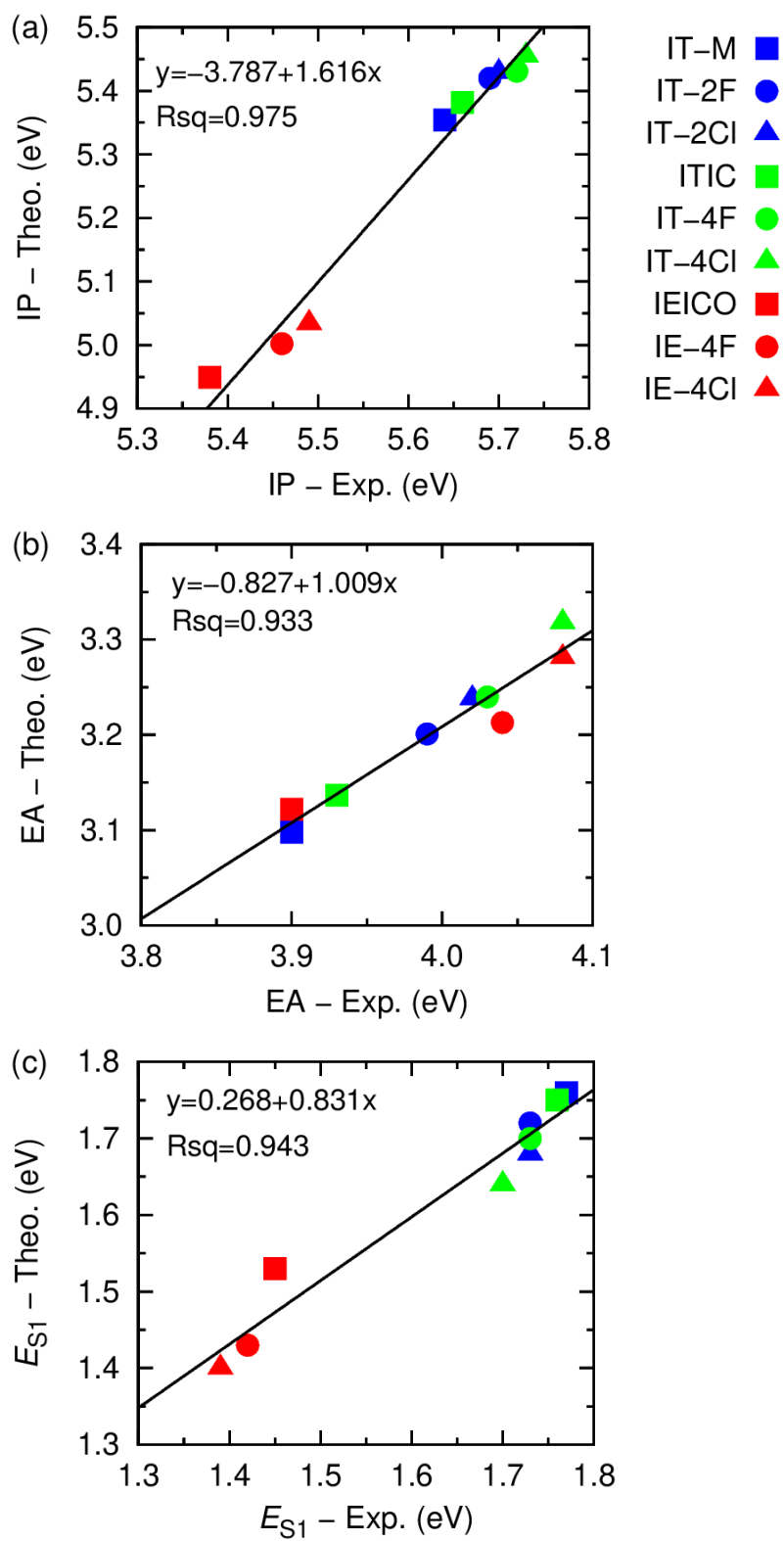

**Figure 2** - Correlation between theoretical and experimental results of IP, EA and $E_{S1}$ for the three groups of NFAs. The linear equation and the the squared correlation coefficient ($R_{sq}$) are displayed in detail.

Although NFAs have a low electric dipole moment due to their symmetry and planarity, when the acceptor units of the molecules are modified by F or Cl substitution, this molecular property undergoes variations induced by the higher electronegativity of those atoms. Indeed, Figure 3a shows that the unsubstituted molecules of the three groups (IT-M, ITIC and IEICO) have an electrical dipole moment very close to zero. With the addition of F atoms in the chemical structure of the molecules the electric dipole moment undergoes a slight increase. For the molecules with Cl the increase is even greater. This effect is related to the strong electronegativity of the F and Cl atoms, making the acceptor groups of the molecules more negative.[44,71] Consequently, the central donor unit becomes more positive. This change in the distribution of the molecular charge can be estimated by obtaining the magnitude of the ICT between the acceptor and donor moieties. The magnitude of ICT for each molecule are presented in Fig. 3b. There is a certain correlation between ICT and the intensity of electric dipole moment of the molecules so that the NFAs with higher ICT within each group tend to have a stronger dipole moment. Note that the molecules of group III have the highest values of ICT. For example, IE-4Cl had ICT = 0.61 electron compared to ICT = 0.51 electron for IT-2Cl (that belongs to group I).

It is known that the electronegativity of fluorine is stronger than chlorine (Pauline electronegativity[72] for Cl: 3.16; for F: 3.98). Yet Figure 3 demonstrates that the acceptor molecules with chlorine presented the highest values of dipole moment and ICT. This result can be related to the significantly longer C–Cl bond (~1.74 Å) than C–F bond (~1.34 Å) that helps to separate the electron from the central core (D unit). These changes in the molecular structure and charge distribution reflect on the electronic energy levels and exciton binding energy. A deeper discussion on the subtle differences on the effects induced by the F or Cl substitution will be developed below.

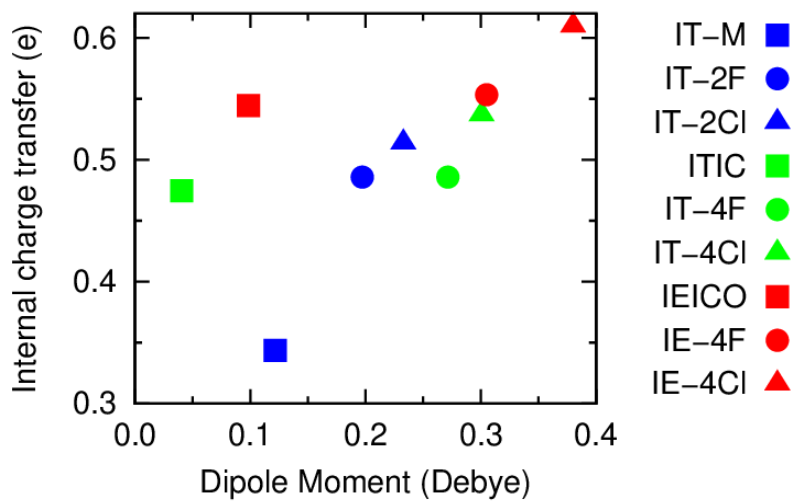

**Figure 3 –** Dipole moment and internal charge transfer for the three groups of NFAs.

Figure 4 shows the calculated vertical excitation energies and oscillator strength (*f*) of the $S_0 \rightarrow S_1$ transition. The magnitudes of *f* indicates a high probability of photon absorption for energies corresponding to the $S_0 \rightarrow S_1$ transition. On the other hand, as expected, *f* = 0 for the $S_0 \rightarrow T_n$ transition so that the direct photoexcitation of triplet states is extremely improbable. Consequently, the triplet generation depends on an efficient intersystem crossing process that can be improved by the reduction of $\Delta E_{ST}$.[35] Another important result from Fig. 4 is that the energies of the excited states $S_1$, $T_1$, $T_2$ and $T_3$ decreased with the F or Cl substitutions.

From Figure 4 it is possible to obtain the energy gap between $T_1$ and $S_1$ states, $\Delta E_{ST}$ (see inset). The fluorination dropped $\Delta E_{ST}$ by 0.01 eV on average. Using DFT calculation, Han and coauthors[30] recently found the same decrease of $\Delta E_{ST}$ with the fluorination of ITIC. Here we obtained that chlorination decreases $\Delta E_{ST}$ by 0.02 eV on average, double of the reduction produced by fluorination.

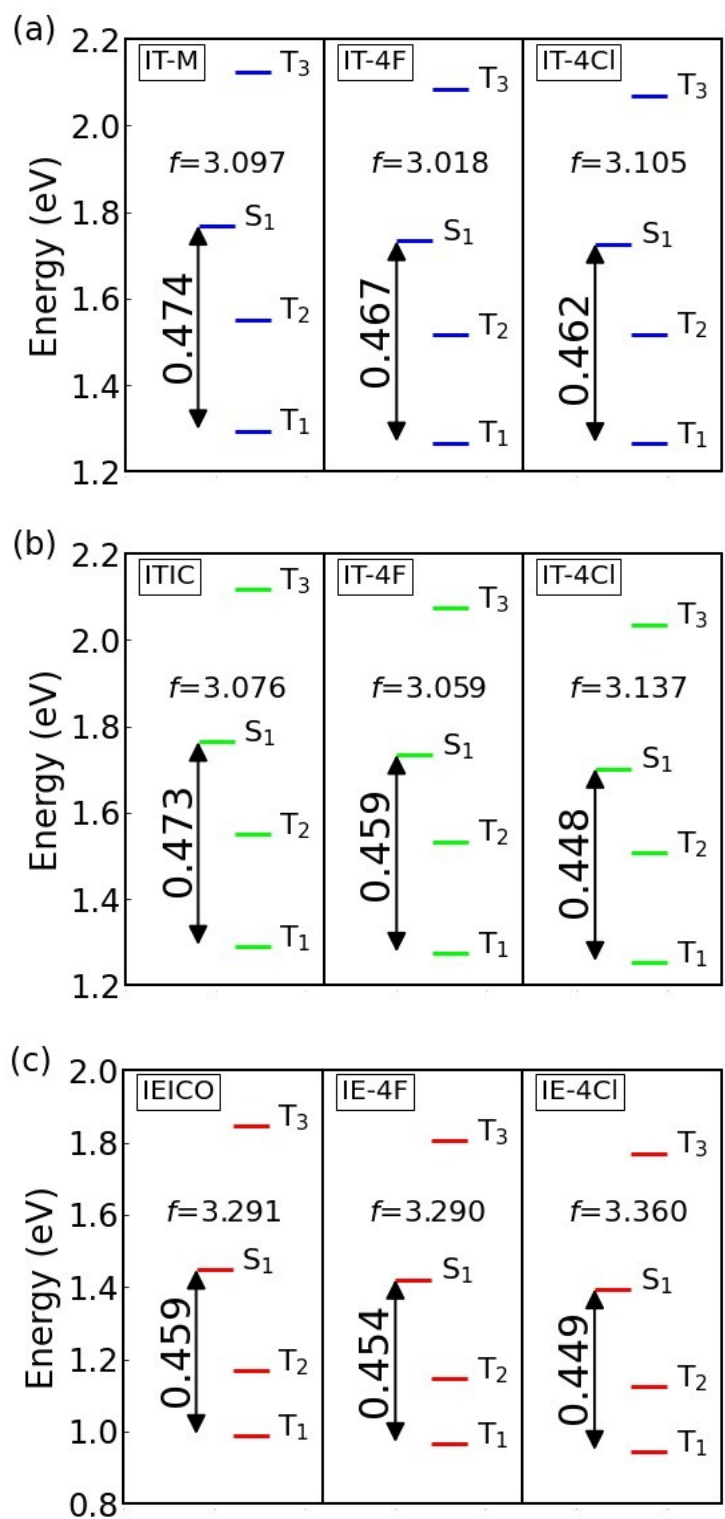

**Figure 4** - Vertical excitation energies and oscillator strength ($f$) of the $S_0 \rightarrow S_1$ transition for the three groups of NFAs. Inset: double arrows represent the energy gap between $T_1$ and $S_1$ states ($\Delta E_{ST}=E_{S1}-E_{T1}$).

The frontier molecular orbitals HOMO and LUMO of the three groups of NFAs was presented in Figure 5 together with the degree of HOMO-LUMO overlap, $\Theta_{H-L}$, and the weights of the HOMO→LUMO transition in the S1 and T1 excitations. Note that the HOMO is localized mainly in the central donor unit of the molecules whereas the LUMO is localized mainly in the end-group. $\Theta_{H-L}$ decreases by approximately 1% with fluorination. The molecules with Cl (apart from group I molecules) have even smaller $\Theta_{H-L}$ compared to molecules with F. From Figs 4 and 5, one can see that there is a relationship between $\Delta E_{ST}$ and $\Theta_{H-L}$ where smaller values of $\Theta_{H-L}$ tend to reduce $\Delta E_{ST}$. This effect is a result of the lower degree of HOMO-LUMO overlap which increases the average distance between the photo-excited electron and hole pair. Since exchange interactions have a short range, this effect approaches the triplet and singlet energies. Furthermore, the IE-4Cl acceptor has the lowest values of $\Theta_{H-L}$ = 62.8%. The extended central unity of IEICO based-molecules helps to decrease the degree of HOMO-LUMO overlap. From our calculations, the molecular length, *l*, of the molecules are presented in Figure 5.

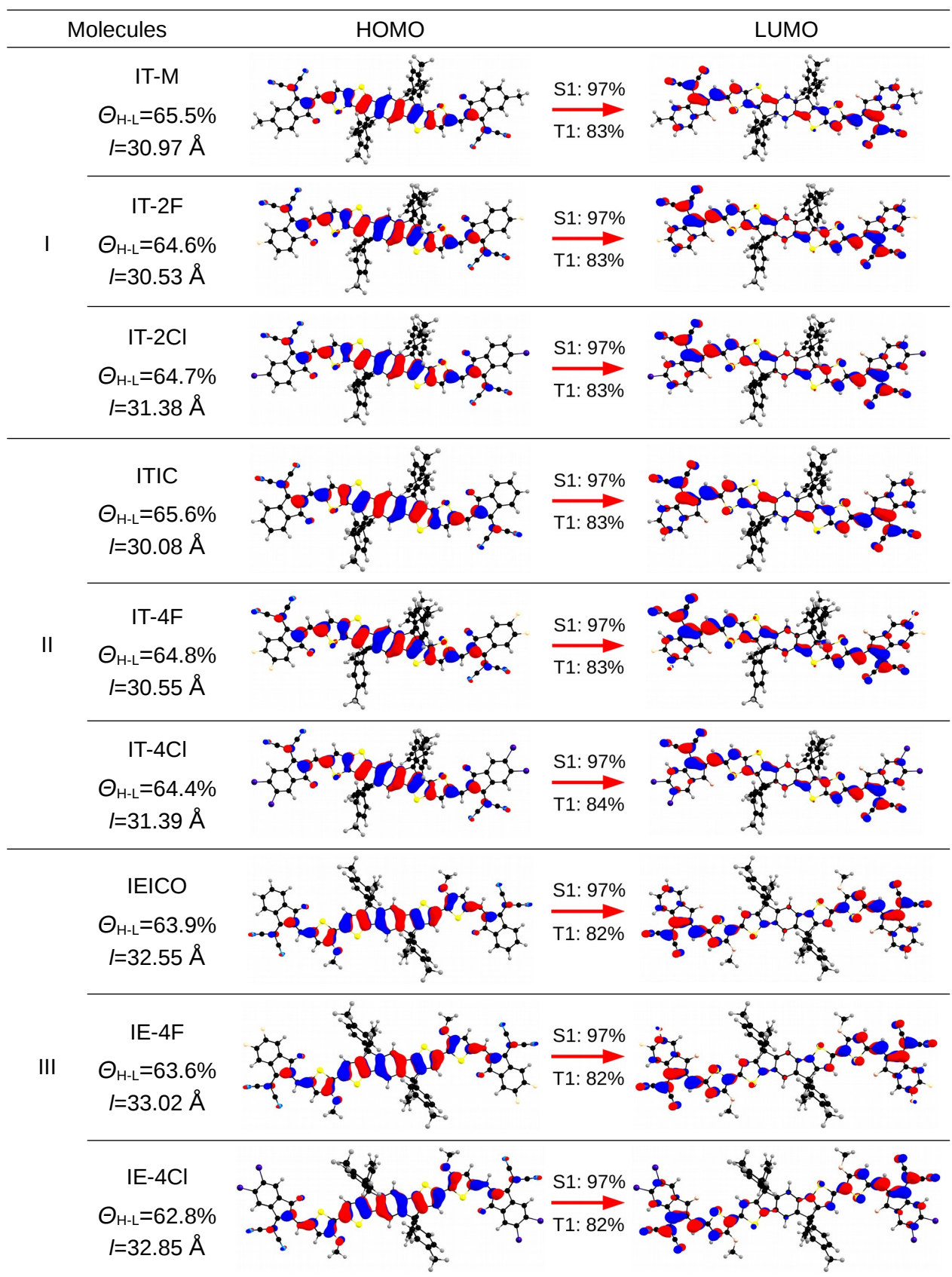

**Figure 5 –** Frontier molecular orbitals for the three groups of NFAs (isovalues 0.02), degree of HOMO-

LUMO overlap, $\Theta_{H-L}$, and molecular length, $l$. The weights of the HOMO→LUMO transition in the S1 and T1 excitations are provided.

Associated to the $\Delta E_{ST}$ reduction, there is consequent decrease of $E_{b,T}$ (Fig. 6). The average drop of $E_{b,T}$ upon fluorination (chlorination) in Fig. 6 was 0.02 eV (0.04 eV). This average decrease is higher compared to the reduction observed for the binding energy of singlet excitons ($E_{b,S}$), around 0.01 eV (0.03 eV) upon fluorination (chlorination). Among all groups, lower values of $E_{b,T}$ and $E_{b,S}$ were calculated for group III NFAs which can be attributed to the longer molecular length of the IEICO derivatives.[73] Within group III, however, IE-4Cl have the lowest values of $E_{b,T}$ and $E_{b,S}$ (0.81 eV and 0.36 eV, respectively). Note that in general $E_{b,T}$ is approximately twice as large as $E_{b,S}$. Although the $E_{b,T}$ magnitudes are high (between 0.96-0.8 eV), these values are similar to the $E_{b,S}$ values of FAs.[19,20] Therefore, it possible to find a suitable combination of D/NFAs blends with a driving force high enough to dissociate triplet excitons generated in the NFAs. Yet an even larger reduction of $E_{b,T}$ is desirable in order to use these excitons also in low driving force systems. In a broader context regarding interfacial energy levels in D/NFAs blends of organic solar cells, it is very desirable that $E_{T1}$ becomes higher than the energy of charge transfer (CT) state or at least close to $E_{CT}$.[74] A higher value of $E_{T1}$ tends to decrease the transfer rate from CT to $T_1$ and increases the transfer rate from $T_1$ to CT. Importantly, this effect would mitigate the non-radiative recombination pathway from $T_1$ to $S_0$, contributing to decrease the voltage losses.[75]

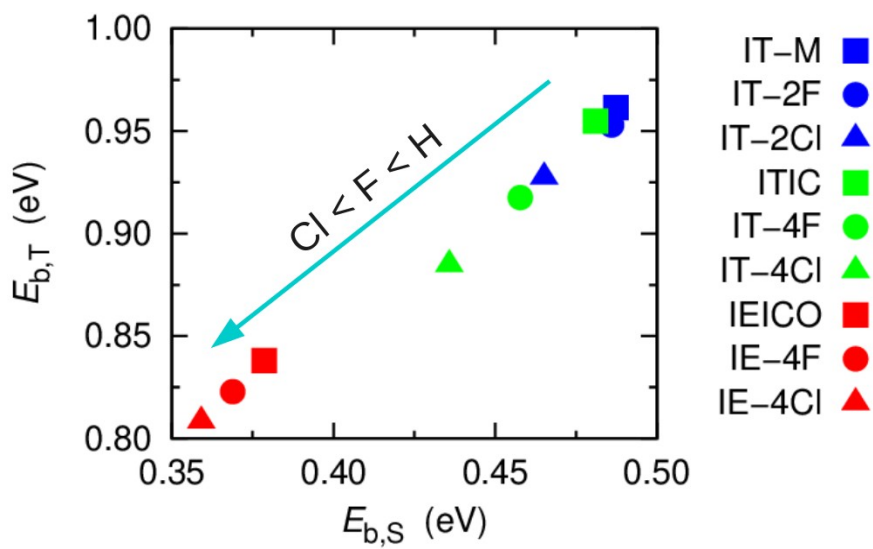

**Figure 6** – Binding energy of triplet exciton and singlet exciton for the three groups of NFAs. The arrow indicates the behavior of energies due to replacement of H atoms by F or Cl.

Even though F is more electronegative than Cl, all the results showed above indicates that the chlorine substitution is more effective than fluorine substitution to decrease $\Theta_{H-L}$ and, consequently, to lower $\Delta E_{ST}$ and $E_{b,T}$. This tendency appears to be counterintuitive and deserves a further analysis. As it was already observed for other organic semiconductors with Cl substitutions, the addition of chlorine atoms tends to significantly stabilize the LUMO but it almost does not change the energy of the HOMO (compared to the unsubstituted or F substitute molecular counterparts[72]). Indeed, this effect can be verified in the results of Table 1 when comparing the EA and IP variations. In addition, Figure 5 reveals differences in the LUMO distribution among Cl and F substitute molecules. There is increased density of LUMO states in the vicinities of the chlorine atoms relative to the density of those states around the F atoms. This result indicates that there will be a higher electronic delocalization once an electron is promoted to unoccupied states of the LUMO for the molecules with Cl substitution. As suggested in Ref.[72], this effect is related to the fact that Cl atoms has empty 3d orbitals to harbor any delocalization of the pi-electrons. In contrast, the next unoccupied atomic orbital of F atoms is the 3s level which is more localized and higher in energy. Moreover, fluorine atoms have a smaller radius that results in considerable electrostatic repulsion in the case of pi-electron delocalization in its vicinities. Hence higher degree of electronic delocalization of the excited electron around the Cl atoms might be the final reason behind the improved properties of the chlorinated NFAs.

At this point it is important to mention that we are using polarizable continuum model to describe the solid-state effects. Since this method is an effective medium approximation, an improvement of the calculations would be obtained by considering intermolecular interaction in different conformations. This approach would reveal the influence of packing in the molecular energies.[76,77] In general, the molecular packing tends to reduce the energies of electronic transitions.[19,31] It was verified by Han *et al*. that this effect is more pronounced for $E_{S1}$ than $E_{T1}$, which further reduces the magnitude of $\Delta E_{ST}$ in approximately 0.1 eV.[30] Additionally, it is also important to investigate the rate of intersystem crossing from singlet to triplet state. This parameter depends on the $\Delta E_{ST}$, the spin-orbit coupling, and the reorganization energy.[78,79]

Finally, in addition to spin-orbit effects, triplet excitons can also be generated from singlet fission through singlet−singlet exciton annihilation (SSA). This effect is especially relevant at high excitation fluences.[33,80,81] Using transient absorption spectroscopy Natsuda *et al*.[82], had recently suggested that this mechanism of triplet formation is dominant in Y6, another important non-fullerene acceptor. Interesting, they attribute the ultrafast triplet exciton formation to higher excited singlet states ($S_n$ states) that satisfies the energetic requirement for singlet fission, $E_S > 2E_T$. It is possible that these mechanisms may also occur with the NFAs studied here and should be investigated in a future work.

In conclusion, we studied the singlet-triplet gap and the binding energy of triplet excitons for nine typical NFAs. Those properties were calculated using the polarizable continuum model with optimally tuned range-separated hybrid functional. We find that the rational fluorination and chlorination of NFAs can be an effective way to improve the internal charge transfer and decrease the degree of HOMO-LUMO overlap of these molecules. These changes favor the generation of triplet excitons and reduce the binding energy of the singlet and triplet excited states specially for those acceptors with Cl substitutions. Among the three groups of NFAs considered here, IEICO derivatives have the lower exciton binding energies due to the longer conjugation length of those molecules. The molecules with the lowest values of $\Delta E_{ST}$ are IT-4Cl (0.448 eV) and IE-4Cl (0.449 eV). Although the magnitude of $E_{b,T}$ (between 0.8 and 0.96 eV) is approximately twice as large than $E_{b,S}$ for NFAs, the magnitudes are similar to $E_{b,S}$ of traditional fullerene acceptors. To our knowledge, this is the first study indicating that the chlorination can be a promising strategy to decrease the binding energy of triplet exciton in commonly used NFAs. In addition, considering the active layer of OSCs formed by D/NFAs bulk heterojunctions, higher values of $E_{T1}$ induced by Cl substitutions of the acceptor might decrease non-radiative recombination losses produced by the CT deactivation via $T_1$ states.

The utilization of triplet excitons to enhance charge generation is one promising step to further improve performance of OSCs. In this context, our findings indicate that rational chlorination of the acceptors can be an additional advantage towards more efficient OSCs. This observation is significant since chlorinated precursors are more easily available and are cheaper than fluorinated ones.[83]

## Conflicts of interest

The authors declare no conflicts of interest.

## Acknowledgments

This work has been partially supported by the Companhia Paranaense de Energia – COPEL research and technological development program, through the PD 2866-0470/2017 project, regulated by ANEEL. This study was financed in part by the Coordenação de Aperfeiçoamento de Pessoal de Nível Superior-Brasil (CAPES)-Finance Code 001. Research developed with the assistance of CENAPAD-SP (Centro Nacional de Processamento de Alto Desempenho em São Paulo), project UNICAMP/FINEP – MCT and Sistema Nacional de Processamento de Alto Desempenho (SINAPAD/SDUMONT). Special thanks go to CAPES-PrInt-UFPR, CNPq (grant 381113/2021-3) and to LCNano/SisNANO 2.0 (grant 442591/2019-5) for financial support. G. Candiotto gratefully acknowledges FAPERJ Process E-26/200.008/2020 for financial support and Núcleo Avançado de Computação de Alto Desempenho (NACAD/COPPE/UFRJ).

Polymers for Organic Solar Cells. *Macromolecules* **2012**, *45*, 607–632.